\documentclass[runningheads,a4paper]{llncs}

\usepackage{amssymb}
\usepackage{graphicx}
\usepackage[numbers,sort&compress,sectionbib]{natbib}

\usepackage{url}
\urldef{\mailamcg}\path|{michael.lange, g.gorman}@imperial.ac.uk|
\newcommand{\keywords}[1]{\par\addvspace\baselineskip
\noindent\keywordname\enspace\ignorespaces#1}

\usepackage{verbatim}
\usepackage{subfig}
\usepackage{epsfig}
\usepackage{graphicx}

\begin{document}

\mainmatter  

\title{Achieving Efficient Strong Scaling with PETSc using Hybrid MPI/OpenMP Optimisation}

\author{Michael Lange\inst{1}
\thanks{The work presented here was funded by Fujitsu Laboratories of Europe Ltd. 
and the European Commission in FP7 as part of the APOS-EU project (grant agreement 277481).}
\and Gerard Gorman\inst{1}
\and Mich\`ele Weiland\inst{2}
\and Lawrence Mitchell\inst{2}
\and James Southern\inst{3}
}

\institute{Applied Modelling and Computation Group,\\
Imperial College London, London, UK\\
\mailamcg\\
\url{http://amcg.ese.ic.ac.uk}
\and
EPCC, The University of Edinburgh, Edinburgh, UK
\and
Fujitsu Laboratories of Europe Ltd., Hayes, Middlesex, UK
}

\titlerunning{Strong Scaling with Hybrid PETSc}
\authorrunning{Lange et al.}

\toctitle{}
\tocauthor{}
\maketitle

\begin{abstract}

The increasing number of processing elements and decreasing memory to core ratio in modern high-performance platforms 
makes efficient strong scaling a key requirement for numerical algorithms.
In order to achieve efficient scalability on massively parallel systems scientific software must evolve across the entire stack 
to exploit the multiple levels of parallelism exposed in modern architectures. 
In this paper we demonstrate the use of hybrid MPI/OpenMP parallelisation to optimise parallel sparse matrix-vector multiplication in PETSc, 
a widely used scientific library for the scalable solution of partial differential equations. 
Using large matrices generated by Fluidity, an open source CFD application code which uses PETSc as its linear solver engine, 
we evaluate the effect of explicit communication overlap using task-based parallelism 
and show how to further improve performance by explicitly load balancing threads within MPI processes. 
We demonstrate a significant speedup over the pure-MPI mode and efficient strong scaling of sparse matrix-vector multiplication 
on Fujitsu PRIMEHPC FX10 and Cray XE6 systems.

\end{abstract}

\section{Introduction}
\label{sec:introduction}

Recent development in High Performance Computing (HPC) architectures has been driven by a clear trend towards greater numbers of lower power cores and a decreasing memory to core ratio. 
Numerical algorithms and scientific software have to adapt to these changes to efficiently utilise the available memory and network bandwidth. 
Hybrid programming techniques, where shared memory programming is combined with inter-node message passing, can be used to exploit the 
multiple levels of parallelism inherent in modern architectures in order to achieve sustainable scalability on massively parallel systems. 

In this paper we describe the addition of OpenMP thread parallelism to the Portable Extensible Toolkit for Scientific Computation (PETSc)~\cite{petsc-user-ref,petsc-efficient}.
PETSc is a widely used library for the scalable solution of partial differential equations and is often used as a key component of large scientific applications. 
Sparse matrix-vector multiplication (spMVM) is by far the most computationally expensive component of sparse iterative linear solvers~\cite{Schubert11}.
Therefore we focus on optimising spMVM within PETSc using hybrid programming techniques and evaluate strong scalability on large numbers of compute nodes.
We demonstrate that using task-based parallelism to hide communication latency can provide significant speedups over naive OpenMP parallelisation.  
Further, explicit thread-level load balancing can be used to give additional increases in performance, 
resulting in significantly improved scalability over pure-MPI implementations in the strong scaling limit.

The matrices used for benchmarking our implementation are extracted from the open source, general-purpose, 
multi-phase computational fluid dynamics (CFD) code Fluidity~\cite{fluidity_manual_v4}. 
Fluidity solves the Navier-Stokes equations and accompanying field equations on arbitrarily unstructured finite-element meshes. 
It is used in areas including geophysical fluid dynamics, computational fluid dynamics and ocean modelling~\cite{Piggot08}. 

\subsection{Sparse Matrix-Vector Multiplication}
\label{sec:matmult_parallel}

PETSc offers a wide range of high-level components required for linear algebra, such as linear and non-linear solvers as well as preconditioners. 
These are based on a suite of parallel data structures which implement basic vector and matrix operations. 
The most computationally expensive operation for solvers and preconditioners alike is the multiplication of sparse matrices with an input vector.

\begin{figure}
\centering
\subfloat[Diagonal matrix elements are multiplied with the local part of the vector while remote vector elements are gathered.]{
\includegraphics[width=0.45\textwidth]{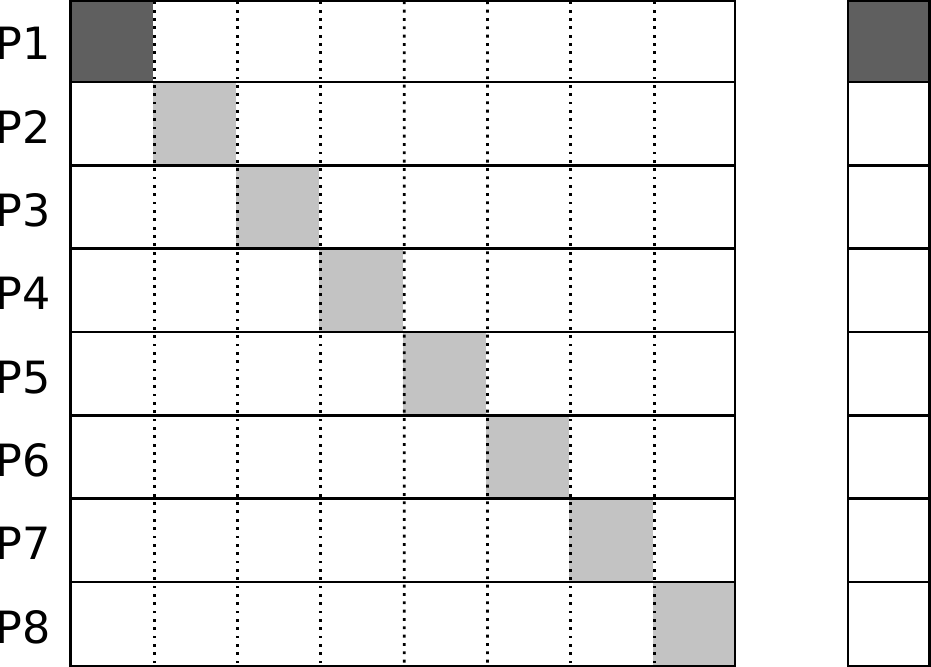}
\label{fig:matmult_local}
}
\hspace{0.04\textwidth}
\subfloat[The off-diagonal sub-matrix is then multiplied with a local copy of the gathered vector elements and added to the partial solution.]{ 
\includegraphics[width=0.45\textwidth]{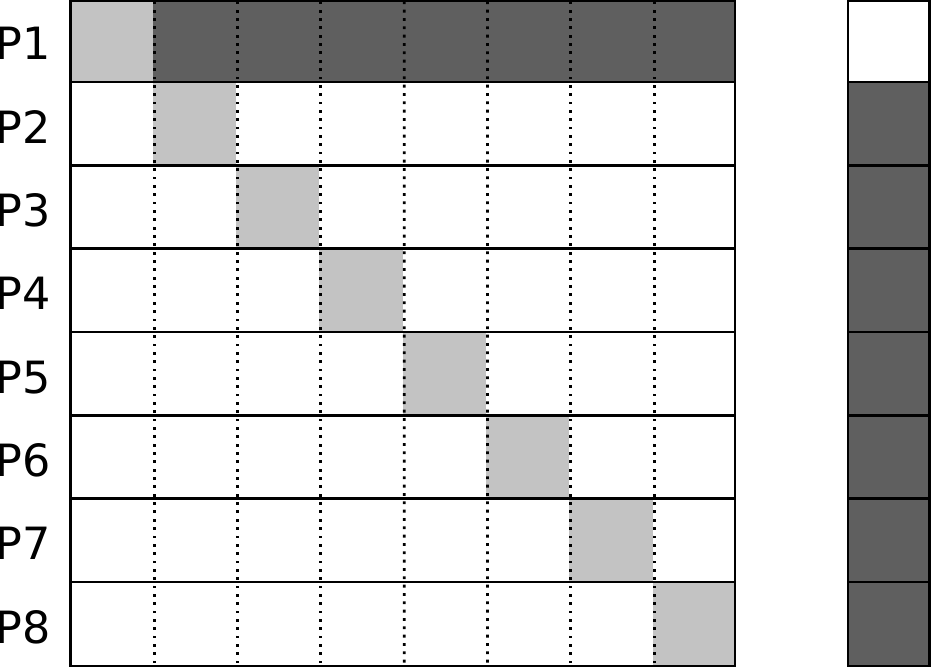}
\label{fig:matmult_offdiag}
}
\caption{Parallel sparse matrix-vector multiplication using 8 MPI processes.}
\label{fig:matmult_parallel}
\end{figure}

PETSc represents distributed MPI matrices by dividing them into diagonal and off-diagonal parts, which on each process are stored as sequential matrices. 
The diagonal sub-matrix hereby corresponds to the part of the input vector that is stored locally by the process.
As a consequence of this storage strategy, as shown in Fig.~\ref{fig:matmult_parallel}, the matrix-vector multiplication is implemented in two phases: 
\begin{itemize}
\item First, each process multiplies its diagonal sub-matrix with the local elements of the input vector, 
while vector elements that reside off-process are gathered into the local memory of the executing process. 
\item Off-diagonal matrix elements are then multiplied with the formerly remote vector elements and added to the partial solution.
\end{itemize}

\subsection{Related Work}
\label{sec:related_work}

Sparse matrix multiplication is one of the most heavily used kernels in scientific computing 
and has therefore received attention from several groups~\cite{Goumas2009,Williams2009,Bell2009,Rabenseifer2009}.
Multiple storage formats, optimisation strategies and even auto-tuning frameworks exist to improve spMVM performance on a wide range of multi-core architectures~\cite{Williams2009}. 
On modern HPC architectures hybrid programming methods are being investigated to better utilise the hierarchical hardware design by 
reducing communication needs, memory consumption and improved load balance~\cite{Rabenseifer2009}.
In particular, task-based threading methods have been highlighted by several researchers, 
where dedicated threads can be used to overlap MPI communication with local work ~\cite{Wellein2003,Rabenseifer2009,Schubert11}.
 
\section{Hybrid MPI/OpenMP Parallelism}
\label{sec:hybrid}

Multi-core processors are now ubiquitous in HPC and programmers are effectively presented with two levels of parallelism: 
inside a compute node, cores share a contiguous memory address space and they can exchange information by directly manipulating this memory space; 
between nodes, distributed memory parallelism is most commonly implemented using explicit message passing via MPI.
Exposing and expressing both intra- and inter-node parallelism can be achieved using a hybrid programming approach.

One motivation for moving away from MPI-only parallelised applications is given by memory limitations. 
While the number of cores is steadily increasing in modern HPC architectures, the memory available to each core is decreasing~\cite{Rabenseifer2009}. 
By exploiting thread-level parallelism, the same number of cores can be utilised within a single node while reducing the MPI memory footprint~\cite{Balaji2009}.
For scientific applications based on domain decomposition, reducing the MPI process granularity also reduces data replication due to halos or ghost cells. 

Performance gains may also be expected from using fewer MPI processes, since it not only reduces communication overheads, 
but also improves the load balance between individual processes~\cite{Schubert11,Rabenseifer2009}.
However, reducing process-level imbalance may have a negative effect on the load balance among threads, 
which in turn can be compensated for by node-level scheduling strategies, as discussed in Sec.~\ref{sec:hybrid_load_balance}.

\subsection{NUMA Architecture}
\label{sec:hybrid_numa}

Non-Uniform Memory Access (NUMA) refers to multiprocessor systems whose memory is divided into multiple memory nodes.
This architecture was designed to overcome the scalability limits of the symmetric multiprocessing (SMP) architecture.
However, this hierarchical memory model for multi-core processors means that it takes longer for a process or thread to access some parts of the memory than others.

It is therefore important to consider data locality in threaded applications, 
since regular off-domain memory access can be detrimental to the performance of already memory-bound applications.
In order to minimise bus contention a parallel \emph{first touch} memory initialisation is often used on NUMA architectures 
to bind data to the memory bank that is closest to the core subsequently using the data block~\cite{Rabenseifer2009}.
In addition, thread and process pinning is required to optimise memory utilisation for all bandwidth-bound algorithms.

When multiplying sparse matrices a master-only approach is most often used to parallelise the local computation steps using threads (see Sec.~\ref{sec:matmult_parallel}).
However, threaded spMVM across multiple NUMA domains requires random but frequent off-domain memory access to fetch input vector elements.
In order to avoid the high-latencies associated with off-domain data fetch NUMA domains can be treated as single address spaces connected by multiple MPI tasks within a compute node.
This approach restricts threads to accessing a single NUMA domain as demonstrated in Sec.~\ref{sec:results_utilisation}.

\subsection{MPI-Communication Overlap}
\label{sec:hybrid_overlap}

As described in Sec.~\ref{sec:matmult_parallel}, PETSc splits parallel spMVM into two phases in order to allow 
the multiplication of the diagonal submatrix to be overlapped with the MPI communication required to fetch off-core vector elements.
Nevertheless, \citet{Schubert11} showed that few MPI implementations provide truly asynchronous communication
and significant performance gains can be achieved by using \emph{task-based} threading, where a single thread is dedicated to actively perform the localisation of global vector elements.
This approach not only overlaps MPI transfer latencies with computation but also hides any sequential overhead incurred from moving data to and from the required MPI buffer space.

\emph{Task-based} threading stands in contrast to traditional \emph{vector-based} threading, where all threads share the computational load evenly.
In order to utilise the \emph{task-based} variant the thread-parallel section needs to be lifted to enclose the vector scatter-gather operation. 
This prohibits the use of OpenMP \verb|parallel for| pragmas to distribute the local row-wise computation among threads 
and requires the explicit computation of thread partition boundaries. 

\subsection{Thread-level Load Balance}
\label{sec:hybrid_load_balance}

Traditional \emph{vector-based} threading with OpenMP divides the number of matrix rows approximately evenly among threads by applying \verb|parallel for| pragmas to the outer loop.
This, however, ignores the fact that individual rows may incur varying amounts of computational work, creating a potential load imbalance within individual thread groups.
Instead, thread-level load balance may be improved statically by dividing the number of non-zeros approximately equally between threads, as pointed out by \citet{Williams2009}.

It is important to note that the matrix stencil does not change during the solve.
Thus, an explicit thread partitioning scheme may be computed after the matrix has been assembled and cached with the matrix object.
This turns the load balance optimisation into a one-off cost, allowing, in principle, the use of load balancing schemes of arbitrary complexity. 

The method used in this paper starts with an initial greedy allocation, where each worker thread receives a block of continuous rows.
This is followed by an iterative local diffusion algorithm, which further balances the number of non-zeros allocated to each thread, 
This procedure balances the thread-level work load and memory bandwidth requirement according to floating point operations required for the solution. 

\section{Benchmark}
\label{sec:benchmark}

The matrices used for benchmarking the hybrid MPI/OpenMP implementations have been generated by Fluidity from a global baroclinic ocean simulation, 
which is representative of a range of three-dimensional multi-scale oceanographic problems~\cite{Piggot08}.
The unstructured mesh is based on two-dimensional high-resolution coastline data that is extruded vertically using constant spacing.
By changing the vertical resolution of the extruded mesh the size of the problem can be scaled linearly, 
allowing a controlled quasi-linear increase in work load for the extracted matrices. 

The benchmark matrices used in this work are pressure field solves extracted after five timesteps. 
The resulting matrices are solved using the Conjugate Gradient method with a Jacobi preconditioner 
and the number of iterations was limited to $10,000$.

\subsection{Cray XE6}
\label{sec:hector}

One of the benchmarking systems used for the work presented here is HECToR, 
a Cray XE6 based on the AMD Opteron 6200 Interlagos processor series and Crays Gemini interconnect~\cite{cray_xe6}.
The Interlagos compute nodes are based on two AMD Bulldozer processors, 
each with 16 cores at $2.3$ GHz paired into two modules and a peak memory bandwidth of $51.2$ GB/s. 
Each module has its own associated memory bank, resulting in four separate memory nodes per compute node~\cite{bulldozer}.

\subsection{Fujitsu PRIMEHPC FX10}
\label{sec:fx10}

The second benchmarking system available to us is a 96-node Fujitsu PRIMEHPC FX10 system~\cite{fujitsu_fx10}. 
The PRIMEHPC FX10 is a UMA (Uniform Memory Access) architecture based on the SPARC64 IXfx processor. 
A single compute node has 16 cores at $1.848$ GHz and a peak memory bandwidth of $85$ GB/s.

\section{Results}
\label{sec:results}

In this section we evaluate the parallel performance of the different hybrid spMVM approaches detailed in Sec.~\ref{sec:hybrid}.
Since hybrid programming offers a complex set of choices on how to utilise a given hardware set, 
we start our investigation by analysing various process-to-thread ratios for fixed numbers of cores. 
This provides insights into the resource utilisation of each algorithm and provides an estimate for the best hybrid configuration 
to be used during the subsequent strong scalability study on large numbers of compute nodes. 

\subsection{Hardware Utilisation}
\label{sec:results_utilisation}

\begin{figure}
\centering
\subfloat[128 cores]{
\includegraphics[width=\textwidth]{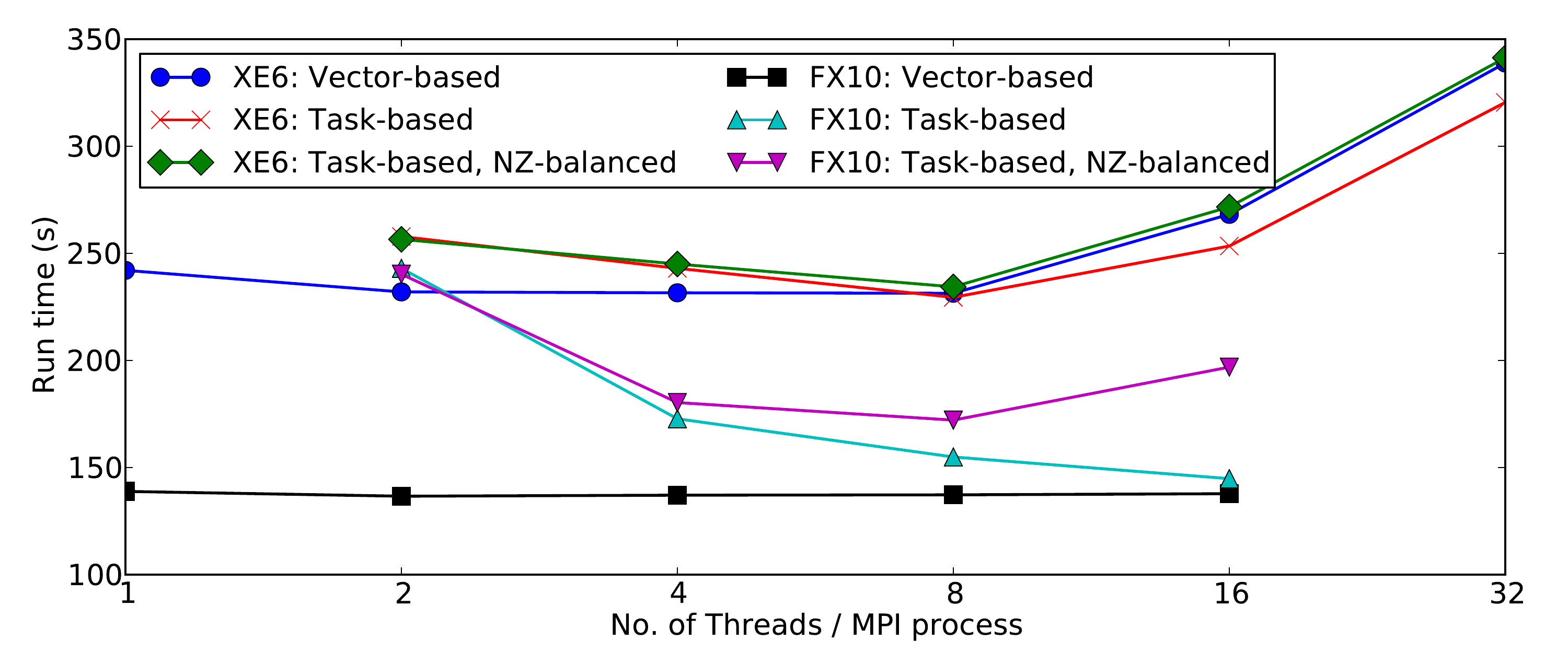}
\label{fig:matmult_nnode4}
}\\
\subfloat[1024 cores]{ 
\includegraphics[width=\textwidth]{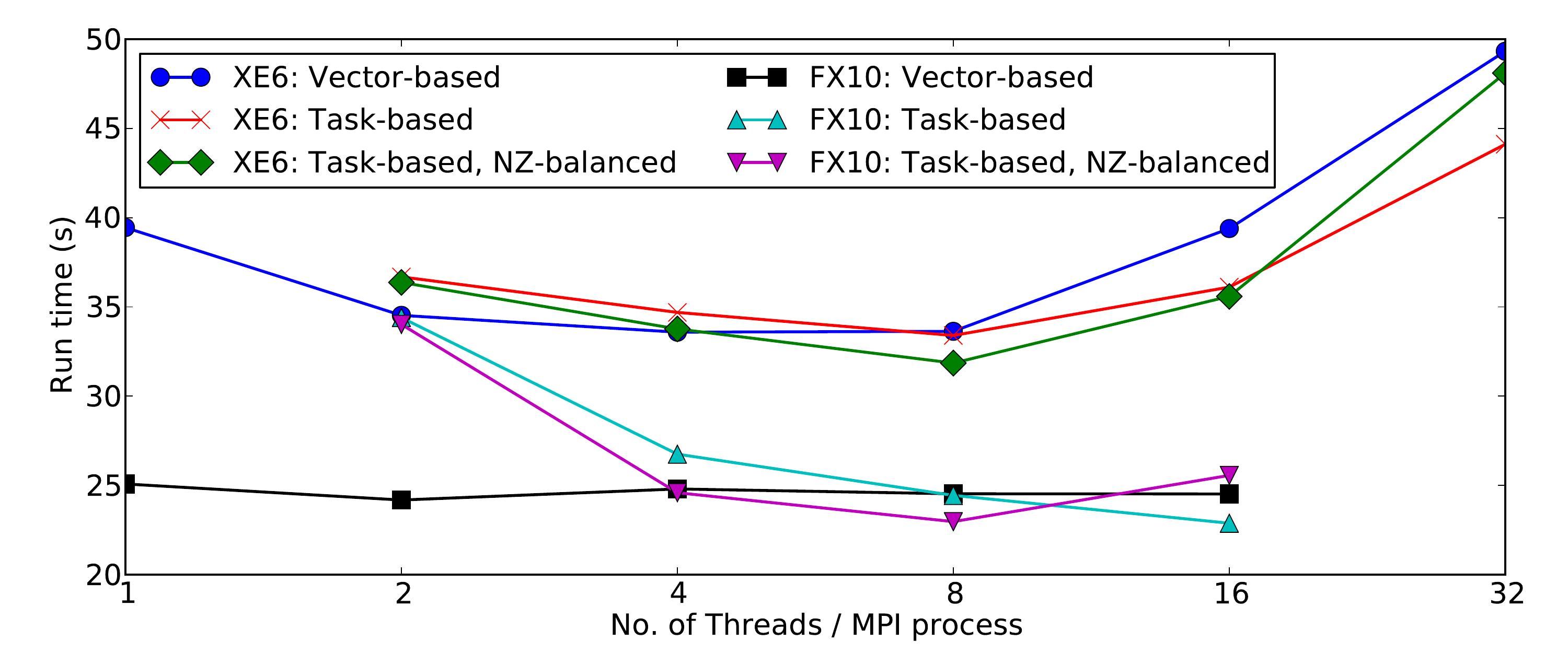}
\label{fig:matmult_nnode32}
}\\
\subfloat[4096 cores]{
\includegraphics[width=\textwidth]{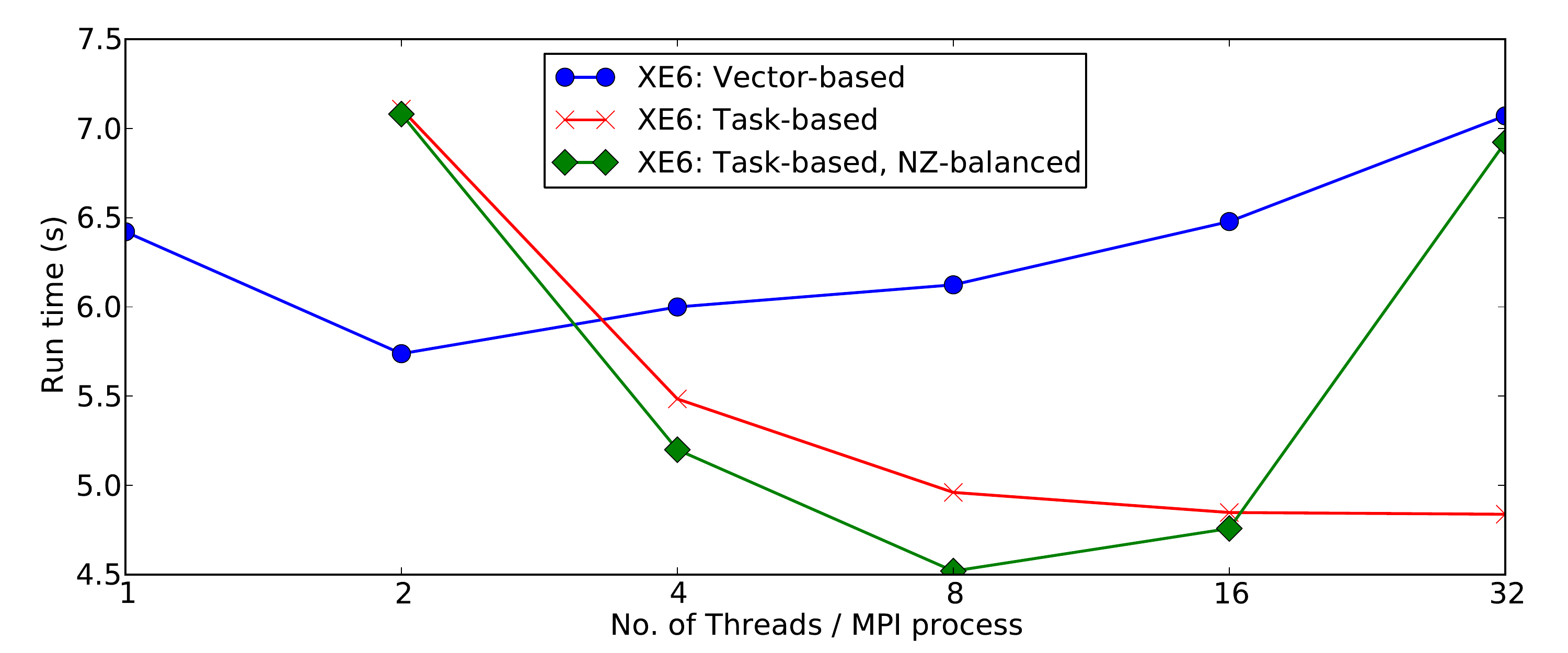}
\label{fig:matmult_nnode128}
}
\caption{Matrix multiplication run times on a fixed number of cores with varying thread-to-process ratios. 
The left most value represents a close approximation to MPI-only performance.
Native compilers were used on both architectures.}
\label{fig:matmult_hybrid}
\end{figure}

Figure~\ref{fig:matmult_hybrid} shows the performance of varying hybrid process-thread combinations on the Cray XE6 and Fujitsu PRIMEHPC FX10 systems. 
The left-most entry of the vector-based configuration constitutes the MPI-only baseline configuration. 
OpenMP overheads have been verified to be negligible for the given problem size using microbenchmarks~\cite{Reid2004}.

On the XE6, using only a small number of compute nodes (Fig.~\ref{fig:matmult_nnode4}~and~\ref{fig:matmult_nnode32}), 
the task-based algorithms with and without explicit thread-level load balancing perform best when running 8 threads wrapped by 4 MPI processes per node. 
This correlates with NUMA alignment, where threads are used only inside individual NUMA domains and MPI tasks connect separate memory nodes. 
A significant performance reduction can then be observed with 16 and 32 threads per process, 
which coincides with NUMA traffic being incurred due to fetching input vector elements (see Sec.~\ref{sec:hybrid_numa}).

However, using 4096 cores (128 XE6 nodes, Fig.~\ref{fig:matmult_nnode128}), the task-based mode without explicit load balancing 
seems to defy the slowdown due to NUMA traffic when using 16 and 32 threads per process.
We can conclude that the algorithm is now bound by memory bandwidth rather than latency. 
In contrast, the thread-balancing mode still experiences a latency slowdown, but exhibits superior performance with a NUMA-aligned configuration.
This is due to an imbalance in vector elements required by each thread due to the explicit thread-balancing, which aggravates the algorithm's sensitivity to memory latency.

Furthermore, both task-based modes significantly outperform the vector-based threading approach on 4096 cores, 
demonstrating the performance loss due to MPI communication overheads. 
Although vector-based threading provides better performance on small numbers of cores due to having an extra worker thread, 
on large numbers of compute nodes the approach struggles to utilise the given memory bandwidth with an increasing number of threads.
As shown in Fig.~\ref{fig:matmult_nnode128}, performance is greatest with only two threads per process, 
indicating that the algorithm's performance is communication-bound.

On the PRIMEHPC FX10 system, we observe similar scaling properties and resource limitations with an increasing number of processing cores for all three algorithms. 
Although the test system used for this work was limited to 1536 cores, we can, therefore, infer an estimate of the the scaling behaviour of the PRIMEHPC FX10 architecture for large scale systems.

The key difference to the XE6 is that PRIMEHPC FX10 is a UMA architecture, and therefore does not incur memory latency penalties due to using multiple memory nodes per thread group.
This can be observed in Fig.~\ref{fig:matmult_nnode4} where, in contrast to the XE6, 
the task-based mode without thread balancing improves performance steadily with increasing numbers of threads per node.
However, the same memory latency limitation on small numbers of cores affects the thread-balancing mode.

On 1024 cores (64 PRIMEHPC FX10 nodes, Fig.~\ref{fig:matmult_nnode32}), the profiles exhibit properties similar to the 4096-core XE6 results. 
The vector-based mode is limited by inter-process communication and performs best with two threads per process, 
while the overall best performance is achieved by the thread-balancing approach using eight threads per process. 

\subsection{Strong Scaling}

In this section we analyse the strong scalability of the described hybrid algorithms on the Cray XE6 system and compare their performance to a pure-MPI approach.
All hybrid modes were run using four MPI processes per compute node with eight threads each in order to prevent NUMA traffic due to input vector elements (see Sec.~\ref{sec:hybrid_numa}).

\begin{figure}
\centering
\subfloat{
\includegraphics[width=\textwidth]{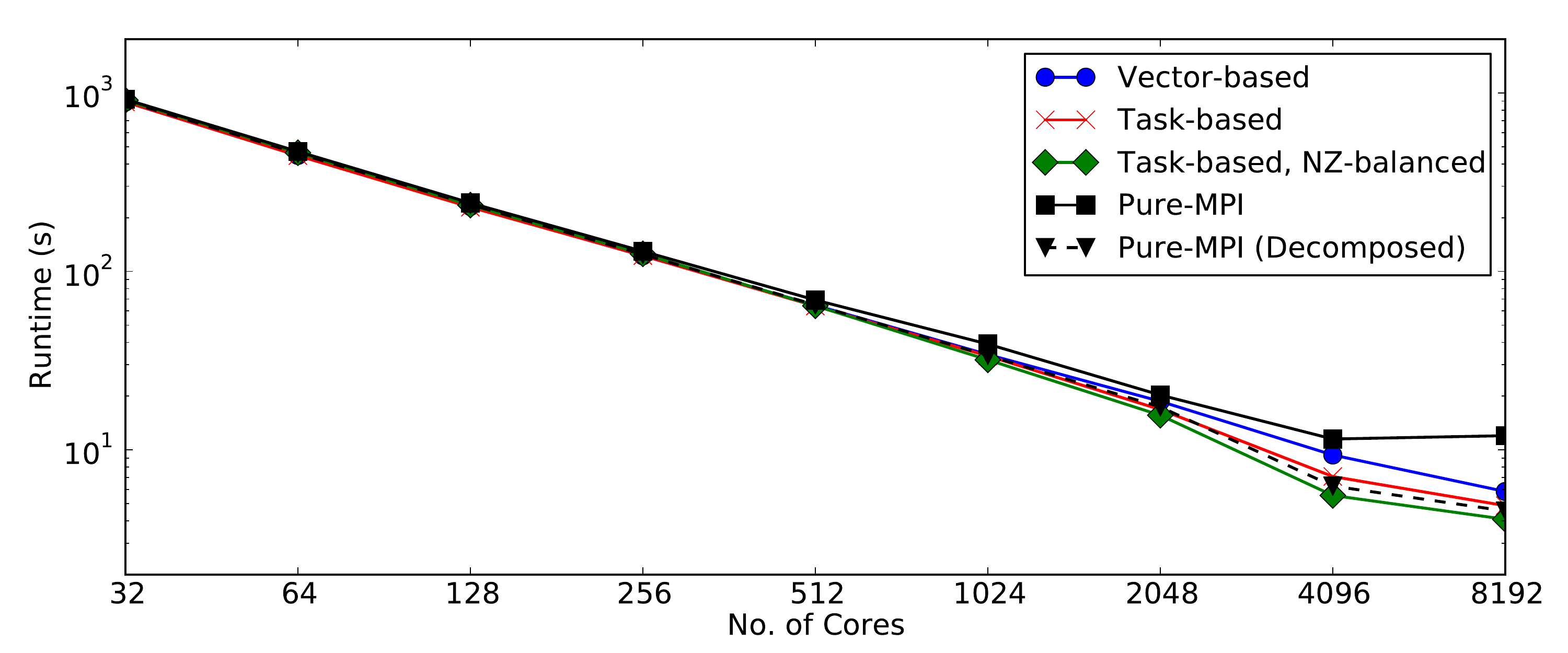}
\label{fig:results_1K_rt}
} \\
\subfloat{ 
\includegraphics[width=\textwidth]{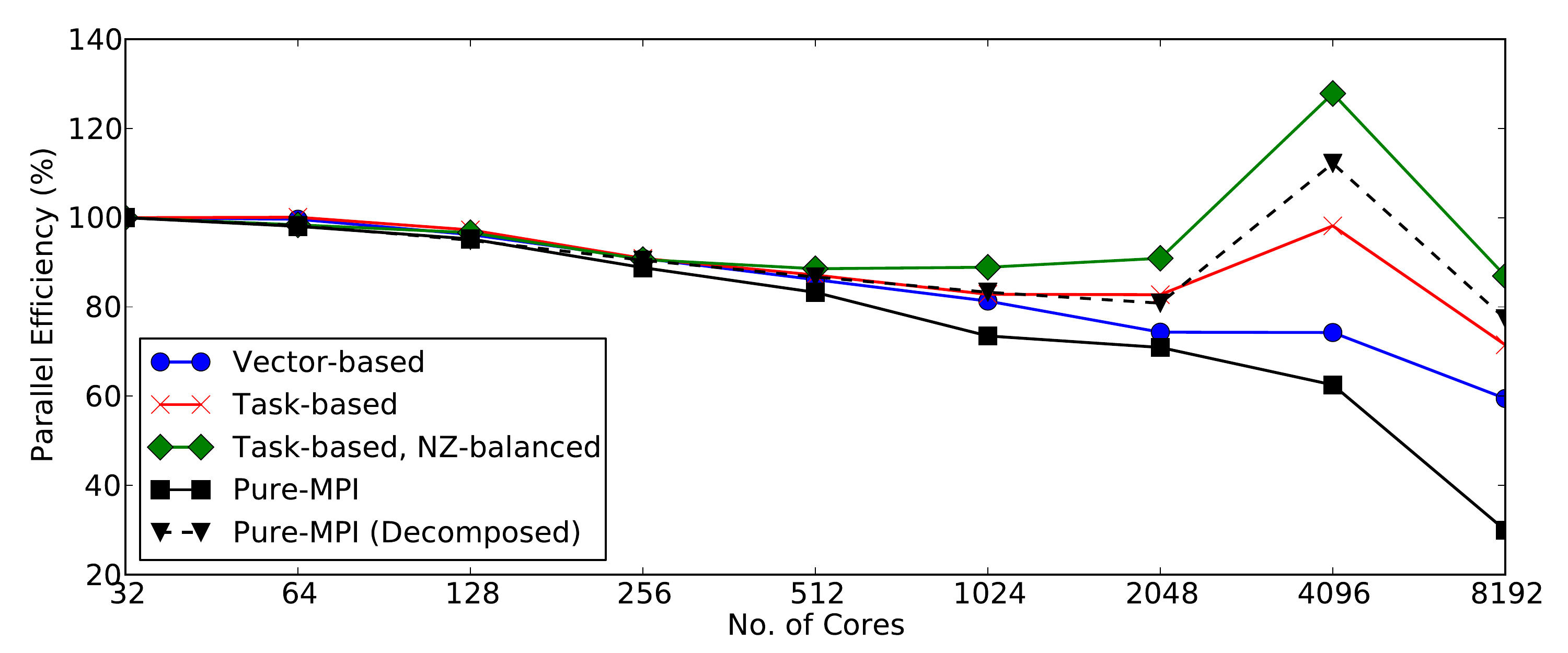}
\label{fig:results_1K_eff}
}
\caption{Strong scaling results for the pressure matrix on up to 256 XE6 nodes (8192 cores). All hybrid modes use 4 MPI ranks per node and 8 threads per rank.}
\label{fig:results_1K}
\end{figure}

The matrix used in Fig.~\ref{fig:results_1K} has 13,491,933 degrees-of-freedom (DoF) and 371,102,769 non-zero elements 
and was generated by a parallel Fluidity simulation decomposed into 1024 sub-domains. 
For the hybrid modes the number of MPI processes used in the strong limit therefore matches the number of processes used during the original decomposition.
For more than 1024 cores, however, the pure-MPI mode uses more processes than the matrix was originally optimised for, 
resulting in a potential slowdown due to load imbalance.
Therefore, an equivalent matrix which has been optimised for 8192 MPI processes has also been included in the benchmark (dashed line).

At the low end of the scaling curve no significant performance differences can be noted.
For more than 512 cores (16 XE6 nodes) the task-based hybrid methods show a better scalability over the vector-based approach.
The thread-balancing implementation hereby performs best, maintaining a nearly constant parallel efficiency of $>88\%$ between 512 and 2048 cores, 
and even experiences slightly super linear scaling between 1024 and 2048 cores. 

On the same matrix, the pure-MPI performance decreases significantly faster than the hybrid algorithms for more than 512 cores (16 XE6 nodes).
The equivalent MPI runs using a more finely decomposed matrix, on the other hand, closely match the performance of the task-based mode without thread-balancing up to 2048 cores.
However, in the strong limit the thread-balancing mode outperforms the optimised MPI runs. 

Furthermore, between 2048 and 4096 cores (64 and 128 XE6 nodes) we observe strong super linear scaling for both task-based methods.
Since the final runtime in the strong limit is below $4$ seconds, we can deduce that scalability ceases at this point due to a lack of computational work 
and that the super linear scaling effects are due to favourable cache effects.

\begin{figure}
\centering
\subfloat{
\includegraphics[width=\textwidth]{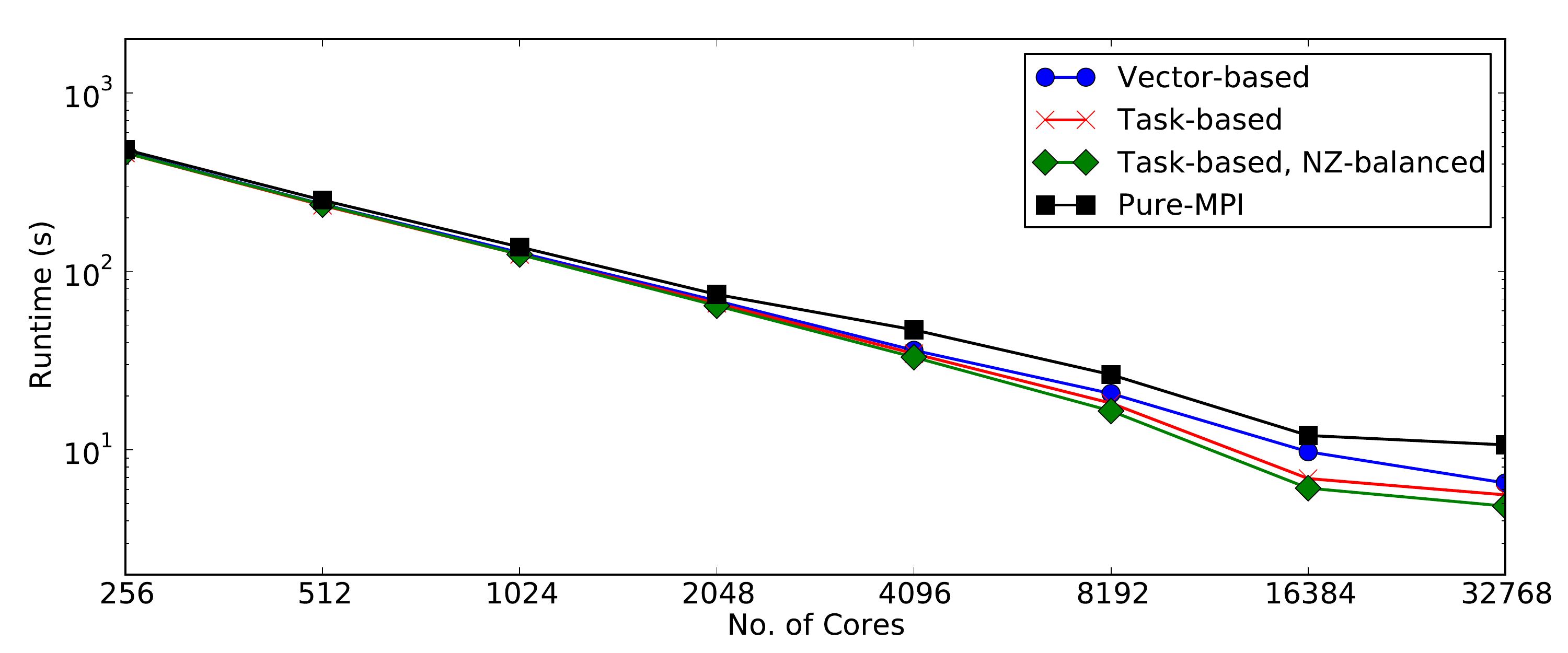}
\label{fig:results_x4_rt}
} \\
\subfloat{ 
\includegraphics[width=\textwidth]{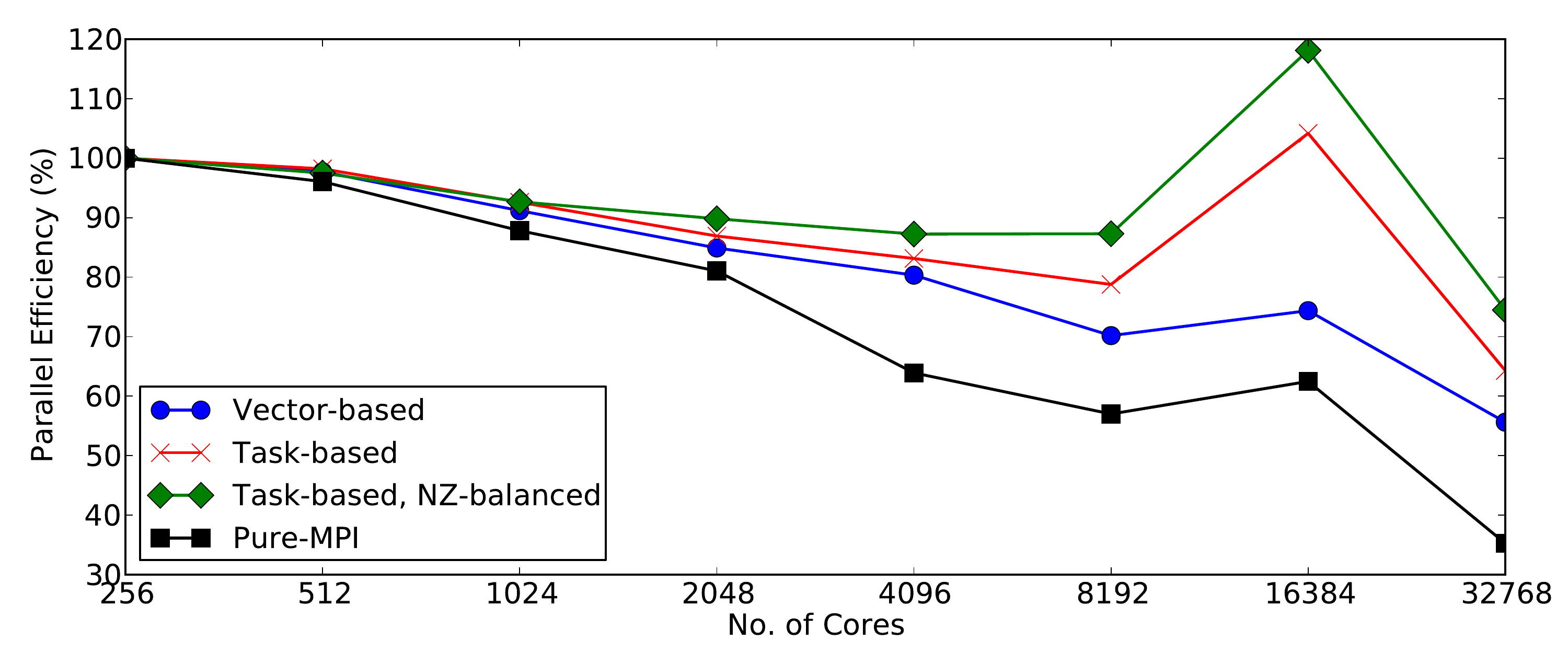}
\label{fig:results_x4_eff}
}
\caption{Strong scaling results for a larger pressure matrix on up to 1024 XE6 nodes (32768 cores). All hybrid modes use 4 MPI ranks per node and 8 threads per rank.
Runs with less than 256 cores (8 XE6 nodes) have been omitted due to insufficient memory per MPI process.}
\label{fig:results_x4}
\end{figure}

Fig.~\ref{fig:results_x4} shows scalability on up to 32,768 cores (1024 XE6 nodes) when the workload of the matrix multiplication is 
increased by a factor of 4 by changing the vertical extrusion of the parent mesh (see Sec.~\ref{sec:benchmark}).
This matrix has 52,040,313 DoF and 1,462,610,289 non-zeros and is based on a 4096-domain partitioning.
The results follow the same general trend, with significant differences in performance observable in the strong end of the scalability curve. 
The pure-MPI performance starts to deteriorate earlier and the super linear scaling in the high end is more pronounced for all approaches. 

\section{Summary and Discussion}
\label{sec:discussion}

In this paper we present an analysis of the scaling properties of sparse matrix-vector multiplication using a hybrid MPI/OpenMP extension to the PETSc library.
We compare hybrid vector-based and task-based algorithms with a pure-MPI variant using large matrices generated by Fluidity.
We describe an extension to the traditional task-based approach, where the load balance among threads is optimised a-priori according to the number of non-zeros in each row.

The thread-balancing extension is shown to give superior performance when scaled to large numbers of compute nodes 
on a Cray XE6 system and on moderate numbers of nodes of a Fujitsu PRIMEHPC FX10 system.
The algorithm achieves this by improving the memory bandwidth utilisation within a given compute node and by hiding MPI communication latency.
This comes at the cost of increased memory latency effects on small numbers of cores, since the algorithm creates an imbalance in input vector elements per thread. 
However, once the main resource limitation of the algorithm shifts to memory bandwidth the thread-balancing approach can improve performance significantly.

Furthermore, the thread-balancing approach enhances one of the fundamental advantages of hybrid programming:
By reducing the number of MPI processes the inherent load imbalance among processes is reduced at the expense of load imbalance among threads.
This is desirable, however, since we can deal with the thread imbalance explicitly by caching an optimised thread partitioning with the matrix. 
As a result, this approach improves work load balance and memory bandwidth utilisation at the compute node level in order to increase overall performance.

\bibliography{bibliography}
\bibliographystyle{splncsnat}

\end{document}